\input amstex
\documentstyle{amsppt}
\pagewidth{35pc}
\pageheight{55pc}
\NoBlackBoxes

\topmatter
\title Exotic Spaces in Quantum Gravity I: Euclidean Quantum Gravity 
in Seven Dimensions
\endtitle
\author Kristin Schleich and Donald Witt
\endauthor
\affil
Department of Physics and Astronomy\\ University of British Columbia\\
6224 Agricultural Road\\ Vancouver, BC V6T 1Z1, Canada
\endaffil

\abstract 
It is well known that in four or more dimensions, there exist 
exotic manifolds; manifolds that are homeomorphic
but not diffeomorphic to each other. More precisely, exotic manifolds 
are the same topological manifold but have inequivalent differentiable 
structures. This situation is in contrast to the uniqueness of the 
differentiable structure on topological manifolds in one, two and 
three dimensions. As exotic manifolds are not diffeomorphic,  one can 
argue that quantum amplitudes for gravity formulated as functional 
integrals should include a sum over not only physically distinct 
geometries and topologies but also inequivalent differentiable 
structures. But can the inclusion of exotic manifolds in such sums 
make a significant contribution to these quantum  amplitudes? This 
paper will demonstrate that it will. Simply connected exotic
Einstein manifolds with positive curvature exist in seven dimensions. 
Their metrics are found numerically; they are shown to have volumes of 
the same order  of magnitude. Their contribution to the
semiclassical evaluation of the partition function for Euclidean 
quantum  gravity in seven dimensions is evaluated and found to be 
nontrivial. Consequently, inequivalent differentiable structures
should be included in the formulation of sums over histories 
for quantum gravity. 
\endabstract

\endtopmatter

\linespacing 1

\TagsOnRight
\NoRunningHeads
\document

\head 1\enspace Introduction \endhead 

The sum over histories formulation of quantum amplitudes is a 
method suited to the study of many issues in field theory, 
especially those involving nonperturbative effects. This approach 
is particularly useful in the study of quantum gravity, 
as it provides a natural connection between the quantum mechanics 
of gravity and its classical limit. Many interesting
results have been formulated and calculated 
using such sums over histories, for example certain proposals
for initial conditions for the wavefunction of the universe 
\cite{1,2} and tunnelling amplitudes for 
pair creation of black holes \cite{3,4,5,6}.

In particular, in Euclidean quantum gravity, 
a history formally consists of a manifold $M^n$ and
a riemannian metric $g$ on that manifold. The partition function, for
example,
is then given by
$$Z = \sum_{(M^n,g)} \exp(-I[g]) \tag 1.1$$ 
where the sum is over physically distinct riemannian histories 
weighted by the euclidean action,
$$I[g]=-{1\over{16\pi G}}\int (R-2\Lambda) d\mu(g). 
\tag 1.2$$
Other quantities such as the wavefunction of the universe can
be similarly constructed in terms of a sum over a suitable
set of physically distinct riemannian histories. Clearly these 
formulations are somewhat heuristic as the many issues 
involved with the concrete specification of these sums 
are not addressed. Nonetheless, these formulations directly include 
the contribution from histories with both different topologies 
and different geometries. Thus they provide
a natural starting point for consideration of many issues in
quantum gravity, especially those involving topology.

Indeed the semiclassical evaluation of such quantum amplitudes
has been used as the starting point for studying
the qualitative effects of quantum gravity and topology change
\cite{see for example Refs. 7,8 as well as the preceding references}. 
This work has produced 
interesting insights into possible large scale effects of quantum
gravity even though it is clear that the results rely on 
known solutions of the theory.  Given this success, 
one is motivated to return to the formal expression for 
the sums over histories in (1.1) and attempt to 
formulate it more carefully. One of the first questions
that naturally arises is, what properties characterize a physically 
distinct history?

This question, as discussed at length in section 2 of \cite{9}, is
a difficult one; a complete specification of the
set of histories and the measure used in expressions such as 
(1.1) is unknown (and perhaps nonexistant for Einstein gravity
as such a specification is equivalent to a quantization of the
theory). Nonetheless, it is useful and possible to address 
this question 
in the context of the histories relevant in semiclassical 
approximation as such expressions should be related to
the low energy limit of a correct quantum theory of
gravity.  It is clear that riemannian metrics $g$ not 
related by  coordinate transformations are physically
distinct. It is also clear that manifolds that have 
different topology are also physically distinct. 
In one, two and three dimensions, these two properties 
suffice to characterize physically distinct histories. However, 
in four or more dimensions there is a third property that
characterizes physically distinct spaces; the specification of 
a differentiable structure on the manifold.
Manifolds with the same topology can have inequivalent 
differentiable structures in four
or more dimensions. Such manifolds are termed
exotic manifolds. The inequivalent differentiable
structures will lead to different properties for physics 
on these manifolds.
Therefore as argued in \cite{9}, not only 
different topologies,
 but also all inequivalent differentiable structures
on each topology should be
considered in the specification of physically 
distinct histories. 

The appearance of exotic manifolds as riemannian
histories for quantum gravity in four or more dimensions 
should not be too surprising. 
Similar arguments for the inclusion of
inequivalent differentiable structures were proposed in 
the context of Kaluza - Klein theories
by Freund \cite{11,12}. The issue of exotic manifolds
has also  been raised by Brans in the
context of classical gravity \cite{13}.
However, a key issue not addressed in these previous 
arguments for the inclusion
of inequivalent differentiable structures is their 
significance: will the 
possibility of having inequivalent differentiable 
structures on a given manifold
make a difference to the results of a calculation?

In this paper we will demonstrate
that it can; the inclusion of exotic manifolds is significant
in the semiclassical limit of functional integrals for
quantum gravity.  We do so by analyzing
 a particular set of exotic manifolds; the Wallach spaces 
 $M_{-42652,61213}$ and $M_{-56788,5227}$. These
manifolds are simply connected homogeneous spaces in 
seven dimensions \cite{14}.
They admit Einstein metrics with positive curvature \cite{15}. 
Furthermore, Kreck and Stolz \cite{16} showed that 
in this family, one can find homeomorphic manifolds that 
are not diffeomorphic; in particular
the Wallach spaces $M_{-42652,61213}$ and $M_{-56788,5227}$.
are a pair of such exotic manifolds. 
 
 We will  numerically solve
the solutions for the Einstein metrics with positive curvature 
in seven dimensions. 
We find that there are actually two such Einstein metrics on 
each space. The
Euclidean actions for these Einstein metrics are the same 
order of magnitude.
These actions on both exotic Wallach spaces are also of the 
same order of magnitude. Finally they are also comparable to
those of Wallach spaces with the same homotopy type.
 As these Einstein metrics are stationary points of (1.2),
they
contribute in semiclassical evaluations of
the partition function for gravity.
Therefore, given that these exotic manifolds contribute in the 
semiclassical limit,
a sum over histories formulation of
 quantum amplitudes for gravity must include a sum over 
 inequivalent differentiable structures in
 addition to those over distinct geometries and topologies. 
  Furthermore, as the Wallach 
spaces are simply connected, restricting
the topologies summed over in (1.1) to
 simply connected spaces as suggested by some authors 
 \cite{see for example Ref. 17 and
 references therein}, cannot remove their contribution.
 Thus the importance of
 inequivalent differentiable structures must not be discounted 
 in the computation of  quantum amplitudes.

 Section 2 provides a summary and discussion of topology and 
 differentiable structure.
 The topology, geometry and differentiable structure of Wallach 
 spaces  are presented in Section 3. Section 4 gives
 the proof of existence of two Einstein metrics with positive 
 curvature on each Wallach space. It then presents the
 numerical solutions for these metrics.
  The contribution of these exotic manifolds to the partition 
 function
 in the semiclassical limit is analyzed in section 5.  Finally,
 a table of numerical solutions for the Einstein metrics
  for the Wallach spaces of of the same homotopy type
  as these exotic manifolds is given in the Appendix.

\head 2\enspace Basics\endhead

In order to discuss the consequences of exotic manifolds 
as riemannian histories in sums over histories for quantum 
gravity, it is  necessary to have
precise definitions. We will begin with summary of 
these definitions and then discuss their meaning and consequences
to this paper.

The topology of a riemannian history is that of a metrizable 
space corresponding to a smooth manifold. A metrizable space is one 
for which open sets can be defined in terms of a distance function. 
A distance function is a real valued symmetric function
for which  given any two points $x,y$ in the space, 
$d(x,y)=d(y,x)\geq 0 $ with the equality holding if and only if 
$x=y$ and  the triangle inequality holds.
Then

\definition{Definition 1} A metrizable space $M^n$ is a smooth 
manifold if
\roster
\item  every point has a neighborhood $U_\alpha$ which is 
homeomorphic to a subset of
${\Bbb R}^n$ via a mapping $\phi_\alpha:U_\alpha \to {\Bbb R}^n$.

\item  Given any two neighborhoods with nonempty intersection, 
the mapping $$\phi_\beta
\phi_\alpha^{-1}:\phi_\alpha(U_\alpha\cap U_\beta) \to 
\phi_\beta(U_\alpha\cap U_\beta)$$
is a smooth mapping between subsets of ${\Bbb R}^n$. 
\endroster
\enddefinition

\noindent
The condition that the space be metrizable is exactly equivalent 
to restricting the space to be paracompact and Hausdorff as 
used in other definitions. A manifold that satisfies 
condition \therosteritem1 only is a {\it topological manifold}. 
A manifold such that the mapping in \therosteritem2 is $C^k$ 
differentiable\footnote{A $C^k$ differentiable map is a continuous
map with  $k$ continuous derivatives.}
 with $k\geq 1$ instead of
smooth is called a {\it $C^k$ differentiable manifold}. 
One such that the mapping in \therosteritem2 is PL homeomorphism
called a {\it PL manifold}.
\footnote{A map $f:M^n\to N^n$ between two manifolds $M^n$ and 
$N^n$ is  PL if each point $p$ in $M^n$ has a cone 
neighborhood $N=pS^{n-1}$ such that  $f(\lambda p + \mu x) = 
\lambda f(p) + \mu f(x)$ where $x$ is in $S^{n-1}$ and
$\lambda,\mu \ge 0$, $\lambda + \mu = 1$. A PL homeomorphism is
a continuous invertible PL map with continuous inverse.}
A 
{\it smooth manifold with boundary}
is one for which condition \therosteritem1 is replaced 
by the requirement
that every point has a neighborhood $U_\alpha$ which is 
homeomorphic to a subset of the upper
half plane ${\Bbb R}^n_+$.
The set $\{(U_\alpha,\phi_\alpha)\}$ is called the {\it atlas} 
of the manifold.
Each element $(U_\alpha,\phi_\alpha)$ in the atlas is called 
a {\it chart}.

A given topological manifold  can have many different atlases. 
For example, one
can produce a different atlas by choosing different mappings 
$\phi'_\alpha$ for
the same neighborhoods. One can also subdivide a neighborhood 
$U_\alpha$ into
new neighborhoods with nonempty intersection and use the restriction 
of $\phi_\alpha$ to each
new neighborhood to define the new charts. Clearly the manifolds 
defined with these atlases
are equivalent. One can more precisely formulate this equivalence 
in terms of homeomorphisms
and diffeomorphisms:

\definition{Definition 2} A homeomorphism is
a continuous invertible map $h:M^n\to N^{n}$ such that its 
inverse $h^{-1}:N^n\to M^{n}$  is also continuous.
\enddefinition

\definition{Definition 3} A diffeomorphism is
a homeomorphism $h:M^n\to N^{n}$ such that $h$ is a differentiable 
invertible map whose  inverse $h^{-1}:N^n\to M^{n}$ is also 
differentiable. 
\enddefinition

\noindent
A homeomorphism characterizes the topological equivalence of 
the two manifolds. A diffeomorphism
also characterizes the topological equivalence, but in addition 
preserves the additional structure
carried in the atlases of the manifolds.  
This additional  structure is physical; inequivalent
differentiable structures will result in different values for 
physical quantities.  Thus the appropriate
notion of physically equivalent manifolds for physics is 
that of a diffeomorphism.

With Definition 3, one can show that a $C^1$ atlas is $C^1$ 
diffeomorphic to a smooth atlas \cite{18}. Therefore, 
questions
regarding the properties of atlases 
can be restricted to a consideration of smooth atlases without 
loss of generality.

The atlas on a manifold provides the information for 
determining the smoothness and differentiability of functions and 
other quantities. The differentiability
of a function is given by the differentiability of its pullback
by $\phi_\alpha$ to each neighborhood $U_\alpha$ in ${\Bbb R}^n$. 
The determination of the differentiability of the pullback of 
the function then proceeds as in ordinary multivariable calculus. 
Consequently, the differentiability of a function will depend not
only on its form but also on the atlas. 

A simple example of how this works is provided by choosing 
different atlases on the real line.  Denote $M=\Bbb R $  with atlas 
consisting of
one chart $(\Bbb R,\phi)$ where $\phi(x) = x$. This first atlas 
is just the identity map to $\Bbb R$.
\footnote{This example is very degenerate. Note
that as all spaces are $\Bbb R$, the 
mappings are functions. Thus we can use the
standard notation for functions on $\Bbb R$ to describe 
them. Furthermore cartesian
coordinates can be used  to explicitly characterize points 
on all copies of $\Bbb R$. We will use
the convention that $x$ corresponds to points in the 
manifold and $y$ to the coordinates in the
chart.} 
Let $N=\Bbb R$ with atlas consisting of 
one chart $\{\Bbb R,\psi\}$ where $\psi(x) = x^{\frac 13}$. 
Observe that both of these atlases are
smooth. There is only one chart for each; thus condition 
\therosteritem2 in Def. 1 is satisfied trivially.

Next consider the function $f:M\to\Bbb R$ given by 
$f(x)=x^{\frac 23}$. This $f$ is not smooth on $M$. 
Recall that the analysis of smoothness
must be carried out in the chart. Now the pullback of 
$f$ to the chart of $M$, $f\cdot\phi^{-1}:\Bbb R \to \Bbb R$, is  
$f\cdot\phi^{-1}(y) = f(y)=y^{\frac 23}$.
This function is continuous, but not everywhere differentiable. 
Thus $f$ is not smooth on $M$.

By the same analysis observe that the function $g:N\to \Bbb R$ given
by $g(x)=x^{\frac 23}$ is smooth. The
pullback of the function to the chart of $N$ is 
$g\cdot\psi^{-1}(y) = g(y^{ 3})=y^2$. It is clear that 
this pullback is smooth.
Thus although $f$ is not a smooth function on $M$, $g$ 
is a smooth function on $N$.

Now although the sets of smooth functions on $\Bbb R$ depend on the 
atlas, the differentiable structure of the real line for 
both atlases is equivalent; $M$ is diffeomorphic to $N$, 
$h:M\to N$ with $h(x) = x^{\frac 13}$. This implies that a 
function that is smooth in $N$ will be diffeomorphic to a 
smooth function in $M$. In the above case, $g$ will correspond to
the smooth function $g\cdot h^{-1}:M\to\Bbb R$ given by 
$g\cdot h^{-1}(x)=g(x^3)=x^2$. Pulling
$g\cdot h^{-1}$ back to the coordinate chart yields
$g\cdot h^{-1}\cdot \phi^{-1}(y)=y^2$, clearly a smooth function. 
Thus although a smooth function with respect to one atlas 
will be diffeomorphic to a smooth function in another
atlas, its form will change.  This is no surprise; a change
of coordinates generically has this effect.

Now it is intuitively obvious that
different atlases on the same topological manifold will result 
in the same physics if they are related by a diffeomorphism. 
However, will all possible choices of
atlases on a given topological manifold
be diffeomorphic to each other? If $M^n$ were homeomorphic to 
$N^n$ but not diffeomorphic,
then $M^n$ and $N^n$  would be inequivalent. This inequivalence 
would not be characterized by their topology but rather carried 
in their atlases. Even so, it would be reflected in
the physics observed on each space.  For example, the set of 
smooth functions on $M^n$ would not correspond
to the set of smooth functions on $N^n$; any homeomorphism 
$h:M^n\to N^n$ will map smooth functions on $M^n$ to continuous
ones on $N^n$. Thus a smooth solution to Laplace's equation
on $M^n$ would only be continuous in $N^n$ and consequently not
be a smooth solution on that space.
 Thus the inequivalence of the differentiable structures of
 $M^n$ and $N^n$ would be
 reflected in the basic properties
of the space such as its geometry and the spectra of 
differential operators such as the Laplacian.

Clearly, the first question is whether this is an issue at all;
are all atlases on a given topological manifold diffeomorphic?
The answer depends on dimension. 
In one, two and three dimensions one can show that all manifolds 
admit a differentiable structure. Furthermore
 any two manifolds that are homeomorphic are
also diffeomorphic \cite{19}. Thus any two atlases on a given 
topological manifold are equivalent
in these dimensions. In four or more dimensions, this is 
no longer the case; manifolds 
that are homeomorphic are not necessarily diffeomorphic:
\definition{Definition 4} If $M^n$ is homeomorphic to $N^{n}$ but not
diffeomorphic, then $M^n$ and $N^{n}$ have 
{\it inequivalent differentiable structures}.
\enddefinition
Manifolds with  inequivalent differentiable structures
are termed {\it exotic manifolds}.

 The result that manifolds admit inequivalent differentiable structures
  was first shown by Milnor \cite{20}. 
 He explicitly constructed exotic 7-spheres
 which are not diffeomorphic to each other.
These exotic 7-spheres $M^7_k$ are constructed
from  3-sphere bundles over $S^4$. The atlas on these spaces
consists of two copies of $\Bbb R^4\times S^3$. The first copy is the 
complement
of the north pole of four-sphere, the second is the complement of 
south pole. Choose as coordinates
 the quaternion pairs $(u,v)$ on each copy with $u$ the quaternionic 
 coordinates on   $\Bbb R^4$.
 Now identify the subsets $(\Bbb R^4 - 0)\times S^3$ of each chart under 
 the diffeomorphism
 $$(u,v)\to (u^\prime,v^\prime) = (u/||u||^2,u^hvu^j/||u||)$$
 where $h$ and $j$ are integers. Intuitively the integers $h$ and $j$ 
 describe how one bundle is twisted
 before it is glued to the other. By calculation of a differentiable 
 characteristic  class $\lambda(M^7)$, Milnor showed
  that for  $(h-j)^2\neq 1 \mod 7$, the manifold has no
 orientation reversing maps. Thus, as $S^7$ with its usual 
 differentiable
 structure has orientation reversing maps,
 $M^7_k$ is  not diffeomorphic to $S^7$.

 Considerable work has been carried out regarding differentiable 
 structures since Milnor's beautiful result and
 much more is now known. In particular, there are topological 
 manifolds 
 that do not admit any differentiable structure in four or more 
 dimensions \cite{21}.
 However, such manifolds do not appear relevant to 
physics as a differentiable structure is required to define derivatives 
of fields and other physical quantities.\footnote{See, for example, 
Appendix B of \cite{10} 
for a summary of how the results of 
Freedman and Donaldson show that the topological
manifold $||E8||$  has only a continuous atlas.}
 In five or more dimensions, the number of 
 inequivalent differentiable structures is determined by an 
invariant $ks(M^n)$ that characterizes whether the structure group 
of the topological manifold Top$(n)$ can be replaced by PL$(n)$:
\proclaim{Theorem 1 [Kirby and Siebenmann \cite{22}]} 
Let $M^n$ be a topological manifold with $n\ge 5$. Then $M^n$
has a PL structure if and only if the invariant 
$ks(M^n)\in H^4(M^n,\partial M^n;\Bbb Z_2)$
satisfies $ks(M^n)=0$. Furthermore, given a continuous homeomorphism 
$h: M^n\to N^n$ between PL manifolds, it is equivalent to a PL
homeomorphism if and only if the invariant 
$ks(h)\in H^3(M^n,\partial M;\Bbb Z_2)$
satisfies $ks(h)=0$.
\endproclaim
 Using these results, it can be proven that there are a finite number
 of inequivalent PL structures on all n-manifolds in five or more 
 dimensions if the cohomology is finitely generated. In particular,
 it implies that as all compact n-manifolds have finitely 
 generated cohomology, they  admit a finite number of inequivalent 
 differentiable structures. 
 
The above theorem classifies PL structures on the manifold. One can 
show that in less than eight dimensions, every PL manifold has a 
smoothing; that is one can smooth the PL structure to obtain a 
differentiable structure. In less than seven dimensions, the 
smoothings are unique. Thus the above theorem classifies 
inequivalent differentiable structures in five and six
dimensions. Furthermore, it can be extended to classify 
the number of inequivalent differentiable structures in seven 
dimensions
using a calculation of characteristic classes that determines the
number of smoothings of a given PL manifold \cite{22}.

 Theorem 1 breaks
 down in four dimensions. The vanishing of $ks(M^n)$ is 
 a necessary condition, but not sufficient in in this case. 
 This is why many issues concerning differentiable structures 
 remain open in four dimensions. However, several
 important facts are known.
First, there are well known examples of 4-manifolds that admit 
more than one inequivalent
differentiable structure. In fact there can be
 a countably infinite number of inequivalent differentiable
structures on compact 4-manifolds. For example,  the connected 
sum of  complex projective space with nine copies of itself 
with opposite orientation, ${\Bbb C}P^2 \# 9(-{\Bbb C}P^2)$, admits a
countably infinite number of inequivalent differentiable 
structures \cite{22}. More surprisingly,
certain open 4-manifolds, notably $\Bbb R^4$, 
admit an uncountably infinite number of inequivalent differentiable 
structures \cite{see for example Ref. 21 and references therein}.
 This result is shocking as all other $\Bbb R^n$  for $n>4$ 
 admit a unique differentiable structure.

From the above, it is clear that exotic manifolds exist 
and are numerous in four dimensions.
 However, it is difficult to directly investigate their 
 consequences to
physics in this dimension. This is because simple 
constructions of exotic 4-manifolds in a form amenable to
further global geometric analysis are unknown.
Therefore it is worth studying the physical contributions of  
higher dimensional exotic manifolds 
as a guide to the possible physical significance of exotic 
manifolds in quantum gravity.

\head 3\enspace Wallach Spaces\endhead

A natural place for the study of physical consequences of 
exotic manifolds
is in quantum gravity. Do these spaces contribute to 
semiclassical evaluations
of functional integrals such as (1.1)? Do to so, 
they must have an 
Einstein metric with positive curvature.
One might be tempted to begin searching for such metrics 
by considering
 exotic 7-spheres.
The manifold $S^7$ with its usual differentiable structure 
clearly admits an
Einstein metric with positive curvature. Some exotic
7-spheres are known not to admit  an Einstein metric with
positive curvature. However, it remains 
an open problem as to  whether or not any of the
exotic 7-spheres also admit such an Einstein
metric \cite{23; see also 11}.
Fortunately, there are known examples of other 7-manifolds 
that
admit both inequivalent differentiable structures and 
Einstein metrics with positive curvature. 
In particular, a set of homogeneous
exotic 7-manifolds admitting Einstein metrics
is provided by the Wallach spaces \cite{24}.
 
 Explicitly,
the Wallach spaces $M_{k,l}$ are coset spaces of the form 
$SU(3)/i_{k,l}(S^1)$
where $i_{k,l}$ is the embedding of $S^1$ in $SU(3)$. 
\footnote{Recall that 
each element $g$ in a Lie group corresponds to a point in 
a manifold. The coset
 space is thus constructed by performing the identification 
 of points in this manifold
 as given by the coset of the group. That is two points $g$ 
 and $g'$ are identified
 if $g=hg'$ for some element $h$ in the subgroup.} 
The generator of $i_{k,l}$ can be written in terms of 
the generators of the standard maximal
torus as\footnote{The maximal torus is the 
 subgroup generated by the maximal
 number of commuting lie algebra generators of the group.}
$${\frak h}_{k,l}=\frac i{\sqrt{2\Gamma_{k,l}}}
\pmatrix k & 0 & 0\\ 0&l&0\\ 0& 0 & -k-l
\endpmatrix \tag 3.1 $$ 
where $\Gamma_{k,l} = k^2 + kl + l^2$. The remaining  
generator is then
$${\frak h}^\perp_{k,l}=\frac i{\sqrt{6\Gamma_{k,l}}}
\pmatrix k+2l & 0 & 0\\ 0&-2k-l&0\\ 0& 0 & k-l
\endpmatrix \tag 3.2$$ 
 that is 
${\frak t} = {\frak h}^\perp_{k,l}\oplus {\frak h}_{k,l}$.
 Insight into the  construction of these spaces can be gained 
 by considering
the action of the exponential
of the generator (3.1) on the subspace of $SU(3)$ 
corresponding
to the maximal torus. This generator identifies points
 along a circle winding around this maximal torus;
$e^{2\pi i\theta} \to  \roman{diag}( e^{2\pi ik\theta},
e^{2\pi il\theta},e^{-2\pi i(k+l)\theta})$.
Different choices of the integers $k,l$ result in different 
points being identified. In
particular, the circle will wind around the maximal torus
a different number of times before closing. Thus the 
homotopy of this circle 
corresponding to restriction of the action of the generator 
of $i_{k,l}$ to the maximal torus depends on 
$k,l$. Of course, this generator also acts nontrivially
on the other elements of lie group. Therefore the 
identification of points on
maximal torus really corresponds to a nontrivial 
identification of a whole set of coordinates
on the manifold.  This identification produces
a particular $S^1$ bundle over the space $M_{k,l}$; that 
which is the  manifold
 $SU(3)$. Now as different choices of the generator $i_{k,l}$ 
 produce different
windings of the circle, it is intuitively plausible that 
they will produce 
$S^1$ bundles over different $M_{k,l}$, all of which are 
isomorphic to  $SU(3)$. Indeed this is the case. 

The tangent space at a point, 
 $T_0 M_{k,l}$, is given by the vector fields associated 
 with the lie algebra
 of the resulting coset space. One can denote these vector 
 fields by their corresponding
 lie algebra elements; 
 $$\alignat3 X^0&= 2\pi\sqrt{6\Gamma_{k,l}} \ 
 {\frak h}^\perp_{k,l}&&\qquad&\\
  X^1&=\pmatrix 0 & 1 & 0\\ -1&0&0\\ 0& 0 & 0
\endpmatrix &\qquad X^3&=\pmatrix 0 & 0 & 1\\ 0&0&0\\ -1& 0 & 0
\endpmatrix &\qquad X^5&=\pmatrix 0 & 0 & 0\\ 0&0&1\\ 0& -1 & 0
\endpmatrix \\
X^2&=\pmatrix 0 & \ i \ & 0\\ \  i\ &0&0\\ 0& 0 & 0
\endpmatrix &\qquad
X^4&=\pmatrix 0 &0 &\  i\ \\0&0&0\\ \  i\ & 0 & 0
\endpmatrix &\qquad
X^6&=\pmatrix 0 &0 & 0\\0&0&\ i\ \\ 0& \ i\  & 0
\endpmatrix .\endalignat$$

The metric for this space can be written in terms of the left 
invariant covectors $\sigma_i$
associated with each $X^i$,
$$ds^2 = \kappa\sigma^2_0+a(\sigma^2_1+\sigma^2_2)
+b(\sigma^2_3+\sigma^2_4)+c(\sigma^2_5+\sigma^2_6).
\tag 3.3$$ 
This space is a homogeneous space for all choices of
 the parameters $\kappa,a,b,c$.
It is thus a seven dimensional
analog of the Bianchi spaces. The
biinvariant metric on $M_{k,l}$ is that naturally induced 
by the group structure
on $SU(3)$  by $-\roman{Re}(\roman{tr}(XY))$. This metric has
coefficients  $\kappa = 24\pi^2\Gamma_{k,l}$, $a=b=c=2$.

From the previous discussion, it is clear that the topology of 
$M_{k,l}$ is 
determined by exactly how the circle is taken out
of $SU(3)$. It turns out that 
several conditions on the integers $k,l$ arise when precisely 
formulating
this removal.
For the case of homogeneous manifolds admitting strictly 
positive 
sectional curvature the action of the subgroup $i_{k,l}$ 
must be nontrivial. Thus
no diagonal term in the generator (3.1) may vanish. 
This restriction is enforced by the requirement 
$kl(k+l)\neq 0$.  For $k,l$ relatively
prime, this identification is on all points for $\theta$ 
in the range 
$0\le \theta\le 1$. Multiplication of both integers by another 
integer $r$ changes
the range of theta to $0\le \theta\le 1/r$ but not the 
identification. Therefore
one can restrict $k,l$ to  relatively prime integers. 

Additionally as
$SU(3)$ contains a $Z_3$ action, integer pairs $k,l$ that are
equivalent mod $3$ result in a group action that contains a 
cyclic subgroup.
For example, the exponential of (3.1) with
 the substitution $k=l+3n$,
$$g_{l+3n,l}(t)=\exp(\frac {ilt}{\sqrt{2\Gamma}}) 
\pmatrix \exp(\frac {3int}{\sqrt{2\Gamma}})& 0 & 0\\ 0&1&0\\ 
0& 0 & \exp(\frac {-3i(n+l) t}{\sqrt{2\Gamma}})
\endpmatrix $$
where $\Gamma=3l^2+12nl+9n^2$ is manifestly the cube of another 
group
element.  Thus
the subgroup $i_{l+3n,l}$ contains
$Z^3$. Therefore  $SU(3)$ will not act effectively on a coset 
space 
$SU(3)/i_{l+3n,l}$ as its action will have fixed points.

Finally, any two coset spaces related by group conjugation 
are equivalent as
group conjugation corresponds to a diffeomorphism on the 
manifold.\footnote{Group
conjugation by an element $g$ is given by the map 
$g\to ghg^{-1}$ applied
to all $h\in SU(3)/i_{k,l}$. If  $SU(3)$ acts effectively, 
this map is invertible, 1-1, onto
and smooth - in other words a diffeomorphism.}
This property allows one to show that $M_{k,l}$ is 
diffeomorphic to 
$M_{-k,-l}$. For example, conjugation  with the element
$$g=\pmatrix 0& -1 & 0 \\ 1 & 0 & 0\\ 0 & 0 & -1 \endpmatrix$$
of elements in $i_{k,l}$ results in the  subgroup 
$i_{l,k}$. This conjugation
of the coset space $SU(3)/i_{k,l}$ will result in 
$SU(3)/i_{l,k}$. Additionally
combining
this group conjugation with the outer automorphism given 
by complex conjugation
will result in $SU(3)/i_{-l,-k}$. This space is also 
diffeomorphic to
$SU(3)/i_{k,l}$. Given these equivalences it is clear 
that one can
restrict the integers $k,l$ to the range $k<0, l>0$ without 
loss of generality.

The Wallach space $M_{k,l}$ is a simply connected manifold 
for all integers 
$k$ and $l$ satisfying the
above restrictions. This can be readily
proven using the fiber bundle exact sequence:
 $$\ldots\pi_1(i_{k,l})\to \pi_1(SU(3))\to\pi_1(M_{k,l})
 \to\pi_0(i_{k,l})\ldots$$
Now $\pi_1(SU(3))=1$ as $SU(3)$ is simply connected; also
$\pi_0(i_{k,l})=1$  as it is connected. Therefore 
the image of the
map to $\pi_1(M_{k,l})$ is the kernel of the map to 
$\pi_0(i_{k,l})$; thus
$\pi_1(M_{k,l})=1$.

Its fourth cohomology is determined from $k,l$:
\proclaim{Theorem 2 (Aloff and Wallach\cite{14})}Suppose 
that $k,l$ are relatively prime. Then
$$H^4(M_{k,l};\Bbb Z) = {\Bbb Z\over \Gamma_{k,l} \Bbb Z}$$
where $\Gamma_{k,l}=k^2+kl+l^2$.
\endproclaim
\noindent Therefore the family of  manifolds $M_{k,l}$ admits 
infinitely many homotopy types.
A characterization of the topology of this family is given by 
Kreck and Stolz. They
derive this characterization in the oriented category; that is 
the manifolds are oriented manifolds
and the diffeomorphisms (and homeomorphisms) are orientation 
preserving.
\proclaim{Theorem 3 (Kreck and Stolz 1991 \cite{16})} 
Assume that both
$k,l$ and $\bar k, \bar l$ are relatively prime. Then
\roster
\item $M_{k,l} $
is homeomorphic to $M_{\bar k,\bar l}$ if and only if 
$\Gamma_{k,l}=\Gamma_{\bar k,\bar l}$ and 
$\lambda_{k,l}\equiv \lambda_{\bar k, \bar l}
\mod 2^3\cdot 3\cdot \Gamma_{k,l}$ where
$\Gamma_{k,l} = k^2 +kl +l^2$ and $\lambda_{k,l}=kl(k+l)$.
\item $M_{k,l} $
is diffeomorphic to $M_{\bar k,\bar l}$ if and only if 
$\Gamma_{k,l}=\Gamma_{\bar k,\bar l}$ and 
$\lambda_{k,l}\equiv \lambda_{\bar k, \bar l}\mod 2^5\cdot 
7^{\mu(\Gamma)}\cdot 3\cdot \Gamma_{k,l}$ where
$\Gamma_{k,l} = k^2 +kl +l^2$, $\lambda_{k,l}=kl(k+l)$ and 
$$\mu(\Gamma) = \cases
0 &\roman{if }\ \ \Gamma \equiv 0 \mod 7\\
1&\roman{otherwise}.
\endcases$$
\endroster
\endproclaim
The diffeomorphism and homeomorphism classifications immediately 
imply 
that certain Wallach spaces with  different values of $k,l$ 
 are diffeomorphic. This follows from
the observation that $\lambda_{k,l}$ is left unchanged by 
certain permutations
of the integers $k,l,(k+l)$. For example 
$$\align 
k,l &\to -(k+l),l \qquad\roman {when}\  k+l>0\\
k,l &\to k,-(k+l) \qquad\roman {when}\  k+l<0.
\tag 3.4\endalign$$
These correspond to orientation preserving diffeomorphisms. 
Other permutations of $k,l,(k+l)$  
change the sign of $\lambda_{k,l}$, for example
$$\align k,l \to -l,(k+l) \qquad\roman {when}\  k+l>0\\
k,l \to (k+l),-k \qquad\roman {when}\  k+l<0.
\tag 3.5\endalign$$
 These
permutations correspond to orientation reversing diffeomorphisms.  
One can explicitly see this
by finding the group element that produces this identification 
under group conjugation. These Wallach
spaces will be distinct in the oriented category as such 
diffeomorphisms are excluded. However, for
the purposes of this paper, one does not wish to consider 
these spaces as physically
distinct manifolds. One can easily augment the 
classification
scheme to handle these cases. As orientation reversing 
diffeomorphisms change the sign of $\lambda_{k,l}$,
manifolds related by an orientation reversing diffeomorphism 
can be classified using
$-\lambda_{k,l}$ in place of $\lambda_{k,l}$ in the 
equivalences in Theorem 3.

This result of Kreck and Stoltz allows one to answer an interesting
question; are there Wallach spaces that
are homeomorphic but not diffeomorphic? The answer is yes; a 
difficult calculation involving
recasting the computation of the modular equivalence in  
Theorem 3 in terms of number theory results
in
\proclaim{Corollary (Kreck and Stolz 1991 \cite{16})} 
If $M_{k,l}$ and $M_{\bar k,\bar l}$ are homeomorphic Wallach spaces
with $\Gamma_{k,l}<2955367597$, then $M_{k,l} $ is diffeomorphic 
to $M_{\bar k,\bar l}$. On the other
hand, the Wallach spaces $M_{-56788,5227}$ and 
$M_{-42652,61213}$ are homeomorphic
but not diffeomorphic. The order of their fourth 
cohomology is $2955367597$.
\endproclaim

Thus the  Wallach spaces with fourth cohomology of 
order $2955367597$
are a natural set to study for understanding the 
contribution of exotic manifolds
to quantum gravity.

\head 4\enspace Einstein Metrics\endhead

An Einstein metric with positive scalar curvature
satisfies $R_{ab}=Eg_{ab}$ for some positive constant $E$. 
All Wallach
spaces with $k,l$  such that $k\neq l \mod 3$ admit an 
Einstein metric \cite{15}. 
However, this method of proof cannot be applied to compute
the metric numerically for a given $k,l$. Thus it is useful 
to have an alternate proof of
this result; it will show that actually  two Einstein metrics 
exist for a given
$k,l$. Furthermore, this derivation can be implemented
numerically to solve for  
the metrics. These will be computed at the end
of this section.

The derivation of the Ricci components
of the metric (3.3) can be carried out using standard 
techniques for coset spaces \cite{24}.
Alternately one can find these components from the metric 
using Cartan calculus to compute
the curvature of the appropriate form of the metric of 
$SU(3)$ and the
Gauss-Codazzi equations to compute the curvature of the 
space orthogonal to the generator of $i_{k,l}$.
Fixing the overall scale of the metric by the condition 
$\kappa = 24\pi^2\Gamma$ where
$\Gamma=\Gamma_{k,l}$, one finds

$$\align
E &= \frac 3\Gamma \biggl(\frac {(k+l)^2}{a^2}+
\frac {l^2}{b^2}+\frac {k^2}{c^2}\biggr) \tag 4.1\\
E &= -  \frac {3(k+l)^2}{\Gamma a^2}+
\frac 4{a}+\frac {a^2-c^2-b^2+2bc}{abc}\\
E &=-  \frac {3l^2}{\Gamma b^2}+
\frac 4{b}+\frac {b^2-a^2-c^2+2ac}{abc}\\
E &=-  \frac {3k^2}{\Gamma c^2}+
\frac 4{c}+\frac {c^2-a^2-b^2+2ab}{abc}\tag 4.2 \endalign $$
 Eliminating $E$ by substitution of (4.1) in (4.2) results in
 $$\align
 \frac 3\Gamma \biggl(\frac {3(k+l)^2}{a^2} +\frac {3l^2}{b^2}+
 \frac {2k^2}{c^2}\biggr) &=\frac {6}{a}
 +\frac {6}{b}-\frac {2c}{ab}\\
 \frac 3\Gamma \biggl(\frac {2(k+l)^2}{a^2} +\frac {3l^2}{b^2}+
 \frac {3k^2}{c^2}\biggr) &=\frac {6}{b}
 +\frac {6}{c}-\frac {2a}{bc}\\
\frac 3\Gamma \biggl( \frac {3(k+l)^2}{a^2} +\frac {2l^2}{b^2}+
\frac {3k^2}{c^2} \biggr) &=\frac {6}{a}
 +\frac {6}{c}-\frac {2b}{ac} \tag 4.3 \endalign$$
Note that these three equations are related
 under simultaneous cyclic permutations of  $a, b, c$ and
 $(k+l),l ,k $. This is a reflection of  the group structure 
 of the Wallach space.
 One can conveniently reexpress this set of equations; 
 defining $$x = \frac ca\ \ \ \ \ \ y = \frac cb \tag 4.4 $$ 
 one finds
 $$c = \frac 3\Gamma \biggl(\frac{3(k+l)^2x^2 +3l^2y^2+2k^2}{
 6x+6y
 -2xy}\biggr) \tag 4.5 $$
where $x$ and $y$ are solutions of 
$$\align
 (6x+6y-2xy)&(2(k+l)^2x^2 +3l^2y^2+3k^2) \\ 
 &\qquad -(3(k+l)^2x^2 +3l^2y^2+2k^2)(6y+6-2\frac yx)
 =0\tag 4.6\\
(6x+6y-2xy)&(3(k+l)^2x^2 +2l^2y^2+3k^2)\\
 &\qquad-(3(k+l)^2x^2 +3l^2y^2+2k^2)(6x+6-2\frac xy)=0. 
 \tag 4.7\endalign$$
Positive  solutions to the above set of equations yield Einstein 
metrics.

These equations inherit a useful symmetry from the Wallach spaces;
equation (4.6) is equivalent to (4.7) under the simultaneous 
exchange of $\{x,y \}\to \{y,x\}$ and
$\{(k+l)^2,l^2\} \to  \{l^2,(k+l)^2\}$. Therefore knowledge 
of the behavior  of one of these equations for
generic coefficients is applicable
to the other. This allows a simple method to show the existence of
a solution. Note that (4.6) and (4.7) are polynomials in 
both $x$ and $y$ of degree less than
five. Thus
one can solve for the roots of each equation for either variable. 
Roots obtained by treating
$y$ as the dependent variable will be parameterized in terms
of $x$ and conversely. Thus these roots generate curves in the 
$xy$ plane. The
simultaneous solutions of these two equations are the points
of intersection of these curves.  Einstein metrics (if there are any)
are at points of intersection with both $x$ and $y$  positive. 
By analyzing these roots, one can prove that there are two
positive intersections of these curves. Therefore there are two 
Einstein metrics on each Wallach space.

We now proceed to do so; observe that (4.6)
 can be rewritten explicitly as a cubic polynomial in $y$ with
$x$ dependent coefficients;
$$\alpha(x)y^3 +\beta(x)y^2
      +    \gamma(x)y          
          +\delta(x)=0 \tag 4.8$$
where
$$\alignat2 \alpha(x)&=6l^2(x^2-1)\\
\beta(x)&= 
   18l^2x(1-x)\\
   \gamma(x)&=4 (k+l)^2x^4 + 6(k+l)^2x^3 + 6 (k^2- (k+l)^2)x^2  - 
          6 k^2 x - 4 k^2\\
          \delta(x)&=-12 (k+l)^2x^4 + 18 (k+l)^2x^3 
          - 18 k^2x^2 + 12 k^2x\tag 4.9\endalignat $$
 Cubic polynomials always have at least one real root.
  Furthermore as the coefficients of (4.8) are polynomials
 in $x$, the roots 
  are piecewise continuous functions of $x$.
The behavior of these roots as curves in the $xy$ plane are 
controlled by the following:
\roster
\item The roots of its polynomial coefficients
in $x$ will provide an upper bound on the number of positive 
roots of (4.6) for all values of $x$.
Those of $\alpha(x)$ and $\beta(x)$ are obvious. The coefficient
$\gamma(x)$ has one nonnegative zero at $x=x_\gamma$ with 
$\gamma(x)<0$ for $x<x_\gamma$
and $\gamma(x)>0$ otherwise; 
$\delta(x)$ has two nonnegative zeros at $x=0$ and $x=x_\delta$ 
with $\delta(x)>0\ \  0<x<x_\delta$
and $\delta(x)<0$ otherwise.  Both $x_\gamma$ and $x_\delta$ 
depend only
on the integers $k,l$. Furthermore, one has that 
$\gamma(1)= \frac 23 \delta(1)$. From this one can
deduce that the nonnegative roots of (4.8) fall into 
either of two cases: 
(1) $x_\gamma<1,x_\delta>1$ corresponding to $l>-2k$ or 
(2) $x_\gamma>1,x_\delta<1$ corresponding
to $l< -2k$.
Now Descartes' rule of signs for the number of solutions of 
(4.8) as
a function of $y$ with coefficients parameterized by $x$ yields\par
{\settabs \+\indent&(1)\quad&$0<x<x_\gamma$\quad&\cr
\+&(i)& $0<x<x_\gamma$ &3 positive roots and 0 negative roots,\cr
\+&& $x_\gamma<x<1$& 1 positive root and 2 negative roots,\cr
\+&&  $1<x<x_\delta$& 2 positive roots and 1 negative root\cr
\+&&  $x_\delta<x$& 3 positive roots and  0 negative roots.\cr
 \+&(ii)& $0<x<x_\delta$& 3 positive roots and 0 negative roots\cr
\+&&  $x_\delta<x<1$& 2 positive roots and 1 negative root\cr
\+&& $1<x<x_\gamma$& 1 positive root and 2 negative roots\cr
 \+&& $x_\gamma<x$& 3 positive roots and 0 negative roots.\cr}
These results set the behavior of the roots as parameterized by $x$. 
In particular, there
is at least one positive root for $0<x<x_{\gamma}$ and 
$x>x_\delta$ ($0<x<x_{\delta}$ and 
$x>x_\gamma$).
\item One can easily solve for $y$ at certain points, namely  
 $\displaystyle y=0,\pm i\sqrt{\frac {2k^2}{3l^2}}$ at  $x=0$ and 
 $\displaystyle y=3,\pm i\sqrt{\frac {2m^2}{3l^2}}x$ as $x\to\infty$. 
 Additionally, one root of
(4.8) at $x=x_\delta$ is $y=0$.  Note that $y=0$ only at 
$x_\delta$ and $0$. Additionally
at $x = 1$, one root is $\displaystyle y= -\frac 32$. As one 
approaches this point, there
are two behaviors for any real solution of (4.8); either 
$y$ approaches  
$\displaystyle y= -\frac 32$ or $y\to \infty$.

\item One can show that $\displaystyle \frac{dy}{dx}=3$ at 
$x=0,y=0$  by taking the derivative
of the equation (4.8). Therefore $y$ is an increasing 
function of $x$ there implying that
there is one positive root in a neighborhood of this point.
\endroster
Fourth, the discriminant of the cubic equation is negative 
except possibly between 
$x_\delta<x<x_\gamma$ for case \therosteritem1 and 
$x_\gamma<x<x_\delta$ for case \therosteritem2.
Therefore there is only one real root outside these 
ranges of $x$. One can deduce from
the results of Descartes' rule of signs that this root is 
positive outside these ranges. Furthermore,
there are no other positive roots for any value of $x$.

Putting these facts together one sees that there are two behaviors 
for the roots of
(4.8) as a function of $x$ depending on the values of  $k,l$.
 For case (1), 
 there is a nonnegative root which is an increasing function 
 of $x$ with initial slope $3$ at 
 the point $x=0,y=0$ 
that increases to infinity as $x\to 1$. For $x>1$
there is a root that crosses the y axis at $x=x_\delta$ and 
increases to $y=3$ as $x\to\infty$.
(This root continues to negative values; at $x=1$ it has value 
$\displaystyle y= -\frac 32$.) This case is
illustrated in Figure 1. For case (2), 
there is a nonnegative root which is an increasing function of 
$x$ with slope $3$ 
at the point $x=0,y=0$  that turns around and crosses the $x$ axis
at $x_\delta$. (It continues to negative values to 
$\displaystyle x=1, y= -\frac 32$.)
For $x>1$ there is a root that is asymptotically decreasing 
from infinity to
a minimum value, then increases to $y=3$ as $x\to\infty$. 
This case is illustrated in Figure 2.

Now, having characterized the behavior of the nonnegative roots 
of (4.6), one has also done so
 for the nonnegative roots of (4.7) as these equations are
 related simply by an exchange of  $\{x,y \}\to \{y,x\}$ and
$\{(k+l)^2,l^2\} \to  \{l^2,(k+l)^2\}$.
Note that under 
 the exchange of parameters,  case (1) behavior may turn into
 case (2) behavior. Therefore there are four possible 
 combinations to analyze when 
searching for intersections of the curves of zeros. However 
it is easy to see that
each combination has exactly two intersections. 

For example, take the case of
the Wallach space $M_{-40388,61811}$ illustrated in Figure 3. 
The
exchange of  $l$ and $(k+l)$ takes a case (1) curve into a 
case 
(2) curve.  In the region of the $xy$ plane for which 
$0<x<1$, the root of (4.6) has  initial slope
 $3$, that of (4.7) has initial slope 
 $\displaystyle \frac 13$. Therefore the second root starts 
 under the first. 
The first root goes to infinity as $x\to 1$; the second crosses
the y axis at finite $y$. It follows that these curves 
must intersect somewhere in this region of the $xy$ plane. 
Therefore there is an Einstein metric with $0<x<1$. 
Next consider the region in the $xy$ plane with $x>1$.
The root of (4.6) is monotonically increasing from $y=0$ 
at $x=x_\delta$ to  $y=3$ as $x\to \infty$. 
The root of (4.7) is always positive. It approaches $y=1$ 
as $x\to \infty$. Also $y\to\infty$
as $x\to 3$. As (4.6) begins from zero and approaches an 
asymptote that is greater than
that of (4.7) as $x\to \infty$, these curves must intersect.
 Therefore there is a second Einstein metric with $x>1$.
 Thus there are two Einstein metrics on each Wallach space.
 
One can similarly analyze each of the remaining three combinations of
roots and show that there are two
intersections in each of these. For example, the 
Wallach space $M_{-56788,5227}$ is illustrated
in Figure 4. The behavior of the roots is slightly different, but 
it is straightforward to
see that again there will be two intersections. Therefore, 
each Wallach space has two Einstein metrics.

Finally, one can show with more work that there can be no more 
than two Einstein metrics
on each Wallach space. Additional positive roots can only occur 
in the regions $0<x<x_\gamma$
and $x_\delta<x$ for case (1) or $0<x<x_\delta$
and $x_\gamma<x$ for case (2). For such a root to occur, one must 
have a double positive
root in this region. Comparison of the form of (4.8) to the 
desired form $(x-x_1)(x-B)^2=0$
where $x_i$ is the known positive root leads to an 
overdetermined set of equations for $B$;
$$\align 3B^2 + 2\frac \beta \alpha B + \frac \gamma \alpha &= 0\\
2B^3 + \frac \beta \alpha B^2 - \frac \delta \alpha &= 0.
\endalign$$
A simultaneous solution to both is needed for the double 
positive root. Necessary
conditions for this to occur are given by the discriminants 
of the above equations. One
finds that
$$\align
\biggl(\frac \beta \alpha\biggr)^2 & > 3\frac \gamma \alpha \\
\biggl(\frac \beta \alpha\biggr)^3 & <27\frac \delta \alpha 
\endalign
$$
One can bound the values of both sides of these two equations 
using the properties of
the coefficients in the regions $0<x<1$
and $1<x$ of the $xy$ plane. From these bounds one can constrain 
the possible values
of $k,l$ for which a simultaneous solution might occur. 
For example, one
finds that for the region $0<x<1$, bounding the coefficients 
$\beta,\gamma,\delta$ 
by their minimum value and $\alpha$ by its maximum value leads 
to the relation $-\frac {16}{17}k <l<-\frac {16}{7}k$. Thus if 
one has a Wallach space
such that both $l<-k $ and $k+l<-k$, this relation will never be 
satisfied and no
double root exists. However, the diffeomorphism equivalence of 
different permutations
of $k,l, (k+l)$ can be used to choose a set of integers $k,l$ 
for each diffeomorphism
equivalent Wallach space that satisfies this constraint. 
Therefore, there are
no additional positive roots to (4.8). Thus the two 
solutions found before
are the only Einstein metrics on  each Wallach space.

The above proof justifies the derivation of the Einstein metrics 
for the Wallach spaces
by numerical methods. The actual values $x,y$ at which the 
curves intersect
 can be solved for using the Newton-Raphson method 
 \cite{see for example Ref. 25}. The
 corresponding values of $a,b,c$ and $E$ can then be calculated 
 using  (4.1), (4.4) and (4.5). 
 This procedure gives metrics with
 different values of $E$ as the derivation of the equations 
 (4.6) and (4.7)
 did not hold $E$ fixed. 
 However, one can find the corresponding Einstein metrics with 
 fixed  curvature by rescaling the metric by
$\lambda$, that is 
$$\kappa,a,b,c\to \lambda\kappa,\lambda a,
\lambda b,\lambda c \tag 4.10$$
 Observe
that under this rescaling,
 $$\align R_{ab} &\to \displaystyle \frac 1\lambda R_{ab}\\
 E&\to \displaystyle \frac E{\lambda^2} .\tag  4.11\endalign$$
 Therefore, one can rescale a metric with a curvature constant 
  of $E$
 to  a curvature constant of $1$ by choosing $\lambda^2=E$. 
 Thus one can find
 the Einstein metrics of constant curvature $1$ by first 
 numerically solving for
 $x,y$, computing the corresponding values of $a,b,c$ and $E$, 
 then rescaling the metric
 using $\lambda=\sqrt{E}$ in (4.10).
 
 Using the above, the metrics and volume for the homeomorphic 
 exotic Wallach spaces\hfill\break
 $M_{-56788,5227}$ and $M_{-42652,61213}$ are summarized in 
 Table 1. The volume is
 given in terms of $\beta$ where
$$V=\beta V_0 \tag 4.12$$
and $V_0$ is the volume of the metric on the Wallach space 
induced by the biinvariant metric on $SU(3)$.
\topinsert\topcaption{Table 1}
\endcaption
\vspace{1pc}
\hskip .35in
\vbox{\topskip=0pt \offinterlineskip
\def\tablerule{\noalign{\hrule}}

\def\mystrut{\vbox to 9.5pt{\vss\vtop to 4.5pt{\vss\hbox to 0pt{}}}}
\def\tablevert{height2pt&\omit&\omit&&\omit&\omit&\omit&&\omit&\cr 
	\tablerule
	height3pt&\omit&\omit&&\omit&\omit&\omit&&\omit&\cr}
	
	\def\drop{height3pt&\omit&\omit&&\omit&\omit&\omit&&\omit&\cr}
	
\halign{ \vrule#\topskip=1em plus 2em\tabskip=16pt&
	\mystrut\hfil$#$\hfil& \hfil$#$\hfil&\vrule#&\hfil$#$
	\hfil& \mystrut\hfil$#$\hfil& 
	\hfil$#$\hfil& \vrule#&\hfil$#$\hfil&\vrule# 
	\tabskip=0pt\cr\tablerule
	height3pt&\omit&\omit&&\omit&\omit&\omit&\omit&\omit&\cr
	&\multispan2\hidewidth \mystrut Wallach Space\hidewidth&&
	\multispan5\hidewidth \mystrut Metric Parameters\hidewidth&\cr 
	height3pt&\omit&\omit&&\omit&\omit&\omit&\omit&
	\omit&\cr\tablerule
	height3pt&\omit&\omit&&\omit&\omit&\omit&&\omit&\cr
	&k&l&&a&b&c&& \hfil \beta \hfil &\cr \tablevert
	
	&-56788	&	5227	&&	1.78458	&	3.71019	
	&	4.66230	&&	12.2181	&\cr \drop
	&\multispan2			&&	5.00997	
	&	4.30654	&	1.91680	&&	15.6404	&\cr \tablevert	
	&-42652	&	61213	&&	2.74990	&	4.01931	
	&	1.60339	&&	7.6311	&\cr	\drop
	&\multispan2			&&	4.87457	
	&	2.03556	&	5.23786	&&	18.9611	&\cr
	height2pt&\omit&\omit&&\omit&\omit&\omit&&\omit&\cr
	\tablerule
	}}
\endinsert
Notice that the volumes for the distinct solutions are
the same order of magnitude; the smallest volume is about 
$40\%$ of
the largest. Furthermore, this difference occurs between 
two Einstein metrics
on the same coset space, $M_{-42652,61213}$. The Einstein 
metrics on  $M_{-56788,5227}$ have volumes
closer to that of the lower action solution.

It is also of interest to compare the properties and magnitudes 
of the volumes of
topologically inequivalent manifolds in the same cohomology 
class as the exotic Wallach spaces. Again these can be calculated
numerically using the Newton-Raphson method.
 The
results of this tabulation
are given in the Appendix.
 This table is ordered by the homeomorphism classification number 
$\lambda_h\mod 2^3\cdot 3\cdot \Gamma$. Some interesting
properties appear. The Wallach space $M_{-62773,31212}$ has 
both extremal volume metrics in this set. 
Both of these metrics
are nearly identical in two of the three parameters. That with 
largest volume has $a\sim b \sim c/2$;
that with smallest volume has $a\sim b \sim 2c$. The manifold 
$M_{-54193,54532}$ has metrics
with nearly equal volumes. For each metric the metric 
coefficients $a,b,c$  differ 
significantly from each other. 
However the two sets of  metric coefficients are, not 
surprisingly, nearly the same.

A different presentation of the data, not given explicitly in this 
paper, reveals an order behind these 
geometric properties.
This presentation is created by relabeling the $M_{k,l}$  
using the  relations
for orientable and nonorientable diffeomorphisms, (3.4) 
and (3.5), as needed
such that each diffeomorphism
inequivalent Wallach space is labeled by the maximal negative 
value of $k$. Now
a sort by increasing $k$ reveals a pattern of monotonic decrease 
of the larger volume metric and a monotonic
increase of the smaller volume metric until  comparable metrics 
are reached. The coefficients $a,b,c$ reveal
a similar pattern of monotonic behavior.
A careful examination of Table 2 indicates 
that the volumes of the two Einstein metrics on the exotic 
Wallach spaces
$M_{-42652,61213}$ and $M_{-56788,5227}$ are in the middle of the 
pack. Therefore, 
there is nothing particularly distinguishing
about the Einstein metrics of these exotic Wallach spaces, 
either from each other or Wallach spaces of the same homotopy type.

\head 5\enspace Semiclassical Evaluation of the Euclidean 
Functional Integral\endhead

Given these numerical solutions for the Einstein metrics 
on the exotic Wallach spaces, one can
evaluate their contribution to semiclassical evaluations of 
Euclidean functional integrals. Clearly,
as the volume of each Einstein metric differs, their 
action will too. However, it is useful to analyze
these effects more quantitatively in the context of a 
simple example. For this purpose, this section will consider
the logarithmic derivative of the partition function 
(1.1);
$$\frac {\partial \ln Z[\Lambda] }{\partial \Lambda}
= \frac 1{8\pi G Z[\Lambda]} \sum_{(M^n,g)}\exp(-I[g]) 
\int d\mu(g) \tag 5.1$$
This quantity is formally the expectation value of the 
volume over all 7-manifolds 
for fixed cosmological constant $\Lambda$.
In semiclassical approximation, the partition function 
becomes a sum over stationary points of
the Einstein action and (5.1) becomes
$$\frac {\partial \ln Z_{sc}[\Lambda] }{\partial \Lambda}
= -\frac 1{Z_{sc}[\Lambda]}\sum_{i} \frac{\partial 
I[g_i]}{\partial \Lambda}\exp(-I[g_i])\tag 5.2
$$
where $i$ indexes the classical Einstein actions corresponding
to the riemannian histories admitting Einstein metrics.
Although (5.2) is formally a sum over all physically 
distinct riemannian histories, one can
also consider quantities defined in terms of sums over 
well defined subsets of these histories. 
In
particular, one can consider the subset to be Wallach spaces 
of a given
topology or  homotopy type. In this case,
$i$ in
(5.2) will index the Einstein actions on the
 subset of histories of interest. 

The classical action for the Wallach spaces can be computed  
using the 
results of section 4 by relating
$E$ to $\Lambda$.
The Euclidean Einstein equation for positive 
cosmological constant in seven dimensions
implies that $R= \frac 75\Lambda$. Therefore 
$E = \frac 15 \Lambda$.
The action (1.2) evaluated for the Einstein metric is then
$$I[g_i]={3V_i\Lambda\over{80\pi G}} \tag 5.3$$ 
where $V_i$ is the volume of the Wallach space 
with Einstein metric $g_i$ with curvature constant
 $E=\frac 15 \Lambda$. 
Using the scaling properties
of (4.11) and (4.12), one can relate this volume 
to $\Lambda$:
 $$I[g_i]= \frac {3V_0}{16\pi G}\biggl(
 \frac 5\Lambda \biggr)^{\frac 34}\beta_i. \tag 5.4$$ 
Note that $V_0$ is the same for all Wallach spaces. 
The  factor $\displaystyle \frac {3V_0}{16\pi G}$
is universal to all Wallach spaces; one can thus choose 
units such that its value is one. Now as
$$\frac{\partial I[g_i]}{\partial \Lambda}=-
\frac 3{20}\biggl(\frac 5\Lambda \biggr)^{\frac 74}\beta_i $$
the semiclassical expectation value of the volume is
$$\align
\frac {\partial \ln Z_{sc}[\Lambda] }{\partial \Lambda} &= 
\frac 3{20} \biggl(\frac 5\Lambda \biggr)^{\frac 74}<\beta>\\
<\beta> &=\frac 1{Z_{sc}[\Lambda]}\sum_{i}  \beta_i\exp(-
\biggl(\frac 5\Lambda \biggr)^{\frac 34}\beta_i)
\tag 5.5 \endalign$$

One can now evaluate (5.5) for sets of Wallach spaces. 
Clearly, the cosmological
constant  will determine the
relative importance of differences in the volume of 
different Einstein metrics. 
First consider the computation of $<\beta>$ for Einstein metrics 
on a single smooth Wallach space,
in particular  $M_{-42652,61213}$ and $M_{-56788,5227}$. 
As $\Lambda \to 0$, the metric with least $\beta$ will 
dominate; thus
$$<\beta(0)>_{-42652,61213}\ =7.6311\ \ 
\ \ <\beta(0)>_{-56788,5227}\ =12.2181$$
In this limit the $M_{-42652,61213}$ has a smaller mean 
$\beta$ as it has the smallest
volume metric.  
As $\Lambda\to\infty$ both metrics on the Wallach space 
will contribute equally; thus
$$<\beta(\infty)>_{-42652,61213}\ =13.2961\ \ 
\ \ <\beta(\infty)>_{-56788,5227}\ =13.9293.$$
The two exotic manifolds have very similar values of $\beta$ 
for this case. 
The above determines the behavior of (5.5) 
for the relevant case of $M_{hom}$ the
topological manifold common to both $M_{-42652,61213}$ 
and $M_{-56788,5227}$. At $\Lambda \to 0$
the contribution from $M_{-42652,61213}$ will dominate, 
but as $\Lambda\to \infty$
all metrics contribute to the result. Thus
 $$<\beta(0)>_{hom}\ =7.6311\ \ \ \ 
 <\beta(\infty)>_{hom}\ =13.6126.
$$
Finally, one can compute $\beta$ for all Wallach spaces 
of the same cohomology from the results tabulated in
the appendix; one
finds
$$<\beta(0)>_{H^4(\Gamma;\Bbb Z)}\ =4.8364\ \ \ 
\ <\beta(\infty)>_{H^4(\Gamma;\Bbb Z)}\ =13.3520.
$$
 The
graphs of the four values of $\beta$ as a function 
of $\Lambda$ are given in Figure 5.

Much
is apparent from this simple calculation. First, there is a 
small difference between
$\beta_{hom}$ and $\beta_{-42652,61213}$ at 
$\Lambda \to \infty$. There 
is a similar difference comparing
the answers on $\beta_{-56788,5227}$ and $\beta_{hom}$ at 
this $\Lambda$. However, for $\Lambda=0$
the story is quite different; $\beta_{hom}(0)=
\beta_{-42652,61213}$ 
but $\beta_{-56788,5227}$
is 60\% greater. Therefore one cannot predict which 
differentiable structure on a topological manifold will provide
the dominant Einstein metric without
carrying out the analysis  for the value of the
cosmological constant of interest.

\head 6\enspace Conclusions\endhead

The examples constructed in this paper show several key features: 
First inequivalent differentiable structures  will 
 contribute on an equal
footing with topology in functional integrals for gravity. 
Secondly, there
is no way to predict a priori which differentiable structure 
if any will
dominate the contribution to the functional integral. 
Furthermore, as the Wallach
spaces are simply connected manifolds,
they will contribute to semiclassical approximations of sums over
histories restricted to simply connected
manifolds. Therefore the contribution of exotic manifolds to 
functional integrals for
gravity must be considered.

Although this paper examined only the Wallach 
spaces of order $2955367597$, 
more examples of
exotic Wallach spaces exist; 30 additional exotic Wallach manifolds
 are known and 14 are tabulated in \cite{16}. 
Observe also that the Wallach spaces are not the only known 
examples of
exotic manifolds that admit Einstein metrics. Kreck and Stoltz 
showed
that the coset spaces 
$SU(3)\times SU(2)\times U(1)/(SU(2)\times U(1)\times U(1))$
in which the subgroup $(SU(2)\times U(1)\times U(1))$ is 
removed in a nontrivial
fashion also contain exotic manifolds \cite{26}. These
 spaces also appear in the physics
literature as manifolds for compactification of supergravity 
theories via
the Kaluza-Klein construction \cite{27}.  
They are not simply connected. However, one
can show that there are topological manifolds in this sequence 
that exhibit all 28 possible inequivalent
differentiable structures. Furthermore,  these exotic manifolds 
admit an infinite
family of Einstein metrics.

A key feature in the construction of the Einstein metrics 
on these Wallach
spaces is their formulation as a coset space. This feature 
yields the remarkably
simple form of their metric (3.3). It is amusing to 
note that due to this,
the differentiable structure for the Wallach spaces is 
encoded in a single analytic
chart. Therefore, although exotic manifolds may seem 
quite mysterious, they may
actually be rather simple once the appropriate 
description is found.

One might be curious as to the implications of such 
exotic manifolds for the 
wavefunction of the universe. One could divide these 
spaces by a six dimensional
manifold and therefore have a stationary point for 
the Einstein action with
boundary. A natural choice for the topology of the 
boundary manifold is $SU(3)/T$,
that is the coset space obtained by removing 
 the maximal torus.
It is known that there are only two distinct homogeneous 
Einstein metrics on
this space: that induced by the submersion of the negative 
of the
 Killing form on $SU(3)$ and a Kahler-Einstein metric. 
 Both of these have two
 equal metric components and cannot come from $M_{k,l}$
  by submersion except
 when $k=l$,or $k=0$ or $l=0$. As the exotic manifolds 
 do not satisfy
 these conditions, they cannot have a
 boundary with a homogeneous Einstein metric. This would 
 seem to imply that
 they do not produce wavefunctions of the universe 
 peaked toward isotropy.
 
 Clearly, an important question is the relevance of these 
 results to four dimensional
Euclidean functional integrals. Unfortunately, one cannot 
use techniques
involving coset spaces similar to those in this paper
to construct examples of exotic 
manifolds in four
dimensions. However, the lack of an easy technique for 
constructing exotic Einstein
 manifolds
by no means implies that these spaces do not exist! 
Furthermore, as
exotic manifolds are in some sense most numerous in four 
dimensions, it could be
argued that their contribution would be most important 
in this dimension.

It is also natural to ask  what these results imply for the 
significance ofdifferentiable structures 
in Lorentzian functional integrals for 
quantum gravity. This is an important question that deserves 
its own forum due to the nature of
the issues involved in formulating such integrals. Indeed 
preparation of such a discussion is a current focus of the 
authors \cite{28}. However, it is perhaps useful to provide here 
a brief discussion of potential significance of inequivalent 
differentiable structures in Lorentzian functional integrals.

Lorentzian functional integrals for gravity involve sums 
over Lorentzian spacetimes; however it not yet clear 
as to what kinds of Lorentzian spacetimes should be included
in these sums. 
For example,  one could formulate Lorentzian functional integrals 
as sums over globally hyperbolic spacetimes  that interpolate
between two spatial hypersurfaces. One could also formulate 
them as sums over more general classes of Lorentzian spacetimes, 
for example a  class allowing regular Lorentzian spacetimes that 
contain closed timelike  curves or one that
allows Lorentzian spacetimes that contain certain "nice" 
singularities. The type of Lorentzian spacetime allowed will 
critically impact  at what points inequivalent differentiable 
structures will contribute.

For example, if one restricts the Lorentzian geometries
to be those corresponding to the evolution of initial data, 
their topology is then that of a n-manifold cross the real 
interval. Furthermore, the  initial value problem
determines the differentiable structure in terms of 
that of the initial Cauchy surface. Thus the 
differentiable structure  of the Lorentzian history is 
determined by that of the initial hypersurface. In
the case of functional integrals in 4 dimensions, the 
spacelike hypersurfaces are 3-manifolds. As all 3-manifolds 
admit a unique differentiable structure, the differentiable 
structure on any evolution is unique. However, for
functional integrals in 5 or more dimensions, the Cauchy
surfaces themselves will potentially admit more than one 
differentiable structure. For these cases, exotic 
differentiable structures will contribute  as part of the 
initial conditions on an equal footing with the topology.

However, one can argue that allowing only  Lorentzian  
geometries   of this type is needlessly restrictive. For 
example, no topology change is allowed for a Lorentzian 
spacetime that is an evolution. If one generalizes the 
Lorentzian  manifolds to a larger class, for example 
those containing closed  timelike curves, then 
inequivalent differentiable structures
may indeed play an important physical role. For example, 
there are an uncountable number of differentiable structures 
on the manifold $S^3\times \Bbb R$ 
\cite{See ref.~21 and references therein}. Smooth 
Lorentzian metrics can be found on each of the inequivalent 
differentiable structures. Therefore, it clearly becomes
an issue whether or not any of these is an appropriate 
Lorentzian solution to the Einstein equations.
 Therefore, the issue of differentiable
structures may have physical significance in  functional 
formulations allowing these classes of Lorentzian 4-manifolds.

Finally, if one allows the use of Euclidean instantons as 
a technique for evaluating Lorentzian functional integrals, 
it is clear that the results on the physical significance 
of inequivalent differentiable structures
immediately apply in four or more dimensions.

\vfill\eject
\head \enspace Appendix A\endhead

The 
Einstein metrics and volumes for the 32 diffeomorphism 
inequivalent Wallach spaces with
$\displaystyle H^4(M_{k,l};Z) = {Z\over \Gamma Z}$ for 
$\Gamma=2955367597$ are summarized in the table
below. The table is ordered by ascending value of the 
homeomorphism classification number
$\lambda_h\mod 2^3\cdot 3\cdot \Gamma$.
\vskip .5truein
\hskip .15truein
\vbox{\topskip=0pt \offinterlineskip
\def\tablerule{\noalign{\hrule}}
\def\mystrut{\vbox to 9.5pt{\vss\vtop to 4.5pt{\vss\hbox to 0pt{}}}}
\def\tablevert{height2pt&\omit&\omit&&\omit&&\omit&\omit&\omit&&\omit&\cr 
	\tablerule
	height3pt&\omit&\omit&&\omit&&\omit&\omit&\omit&&\omit&\cr}
	
	\def\drop{height3pt&\omit&\omit&&\omit&&\omit&\omit&\omit&&\omit&\cr 
	}
	
\halign{ \vrule#\topskip=1em plus 2em\tabskip=10pt&
	\mystrut\hfil$#$\hfil& \hfil$#$\hfil&\vrule#&\hfil$#$\hfil&&\vrule#&\hfil$#$\hfil& \mystrut\hfil$#$\hfil& 
	\hfil$#$\hfil& \vrule#&\hfil$#$\hfil&\vrule#\tabskip=0pt\cr\tablerule
	height3pt&\omit&\omit&\omit&\omit&&\omit&\omit&\omit&\omit&\omit&\cr
	&\multispan4\hidewidth \mystrut Wallach Space\hidewidth&&\multispan5\hidewidth \mystrut Metric Parameters\hidewidth&\cr 
	height3pt&\omit&\omit&\omit&\omit&&\omit&\omit&\omit&\omit&\omit&\cr\tablerule
	height3pt&\omit&\omit&&\omit&&\omit&\omit&\omit&&\omit&\cr
	&k&l&&\lambda_h\mod 2^3\cdot 3\cdot \Gamma&&a&b&c&& \omit\hfil $\beta$ \hfil &\cr \tablevert

&	-40388	&	61811	&&	530781252	&&	2.51431	&	3.85991	&	1.57678	&&	6.73404	&\cr \drop
&\multispan2				&&\omit		&&	4.96699	&	2.05195	&	5.25276	&&	19.4401	&\cr \tablevert

&	-58137	&	49573	&&	1563779580	&&	3.49151	&	1.73778	&	4.52134	&&	11.0571	&\cr \drop
&\multispan2				&&\omit		&&	4.46916	&	5.09034	&	1.95248	&&	16.6112	&\cr \tablevert

&	-57748	&	50187	&&	3724397916	&&	3.55872	&	1.75197	&	4.56521	&&	11.4074	&\cr \drop
&\multispan2				&&\omit		&&	4.42169	&	5.06777	&	1.94214	&&	16.3275	&\cr \tablevert

&	-53223	&	55436	&&	3938960388	&&	3.89576	&	4.77725	&	1.82538	&&	13.2481	&\cr \drop
&\multispan2				&&\omit		&&	4.14816	&	1.88159	&	4.92452	&&	14.7037	&\cr \tablevert

&	-45964	&	60007	&&	5728436076	&&	3.10134	&	4.25978	&	1.66055	&&	9.14854	&\cr \drop
&\multispan2				&&\omit		&&	4.70684	&	2.00273	&	5.189	&&	18.017	&\cr \tablevert

&	-50277	&	57689	&&	6552905052	&&	3.5686	&	4.57161	&	1.75407	&&	11.4594	&\cr \drop
&\multispan2				&&\omit		&&	4.41453	&	1.94057	&	5.0643	&&	16.2848	&\cr \tablevert

&	-49197	&	58364	&&	6923017548	&&	3.45048	&	4.49436	&	1.72922	&&	10.8462	&\cr \drop
&\multispan2				&&\omit		&&	4.49712	&	1.95853	&	5.10324	&&	16.7781	&\cr \tablevert

&	-54193	&	54532	&&	8827144572	&&	4.00552	&	4.84275	&	1.84978	&&	13.8743	&\cr \drop
&\multispan2				&&\omit		&&	4.04418	&	1.8584	&	4.86531	&&	14.0975	&\cr \tablevert

&	-62508	&	26261	&&	9012685956	&&	1.57654	&	2.10501	&	3.60363	&&	5.45457	&\cr \drop
&\multispan2				&&\omit		&&	5.24754	&	5.0972	&	2.07113	&&	20.0076	&\cr \tablevert

&	-40273	&	61837	&&	9036905268	&&	2.50249	&	3.85203	&	1.5758	&&	6.69181	&\cr \drop
&\multispan2			&&\omit		&&	4.97125	&	2.05266	&	5.25315	&&	19.4611	&\cr \tablevert

&	-39492	&	62003	&&	9245608380	&&	2.42268	&	3.79933	&	1.57034	&&	6.41411	&\cr \drop
&\multispan2			&&\omit		&&	4.99919	&	2.05722	&	5.255	&&	19.5955	&\cr \tablevert

&	-61167	&	18364	&&	10381999620	&&	1.60559	&	2.76577	&	4.03016	&&	7.69512	&\cr \drop
&\multispan2				&&\omit		&&	5.23637	&	4.86781	&	2.0343	&&	18.9245	&\cr 
height2pt&\omit&\omit&&\omit&&\omit&\omit&\omit&&\omit&\cr
	\tablerule
	}}

\vfill\eject
\bigskip
\hskip .15truein
\vbox{\topskip=0pt \offinterlineskip
\def\tablerule{\noalign{\hrule}}
\def\mystrut{\vbox to 9.5pt{\vss\vtop to 4.5pt{\vss\hbox to 0pt{}}}}
\def\tablevert{height2pt&\omit&\omit&&\omit&&\omit&\omit&\omit&&\omit&\cr 
	\tablerule
	height3pt&\omit&\omit&&\omit&&\omit&\omit&\omit&&\omit&\cr}
	
	\def\drop{height3pt&\omit&\omit&&\omit&&\omit&\omit&\omit&&\omit&\cr 
	}
	
\halign{ \vrule#\topskip=1em plus 2em\tabskip=10pt&
	\mystrut\hfil$#$\hfil& \hfil$#$\hfil&\vrule#&\hfil$#$\hfil&&\vrule#&\hfil$#$\hfil& \mystrut\hfil$#$\hfil& 
	\hfil$#$\hfil& \vrule#&\hfil$#$\hfil&\vrule#\tabskip=0pt\cr\tablerule
	height3pt&\omit&\omit&\omit&\omit&&\omit&\omit&\omit&\omit&\omit&\cr
	&\multispan4\hidewidth \mystrut Wallach Space\hidewidth&&\multispan5\hidewidth \mystrut Metric Parameters\hidewidth&\cr 
	height3pt&\omit&\omit&\omit&\omit&&\omit&\omit&\omit&\omit&\omit&\cr\tablerule
	height3pt&\omit&\omit&&\omit&&\omit&\omit&\omit&&\omit&\cr
	&k&l&&\lambda_h\mod 2^3\cdot 3\cdot \Gamma&&a&b&c&& \omit\hfil $\beta$ \hfil &\cr \tablevert

&	-20148	&	61561	&&	13361497092	&&	1.58721	&	3.93119	&	2.62036	&&	7.12507	&\cr \drop
&\multispan2			&&\omit		&&	5.24771	&	2.04509	&	4.92718	&&	19.2389	&\cr \tablevert

&	-62773	&	31212	&&	13384952580	&&	1.72257	&	1.74183	&	3.4643	&&	4.8364	&\cr \drop
&\multispan2				&&\omit		&&	5.19875	&	5.1935	&	2.07845	&&	20.226	&\cr \tablevert

&	-48011	&	59028	&&	13422449916	&&	3.32168	&	4.40874	&	1.70292	&&	10.1993	&\cr \drop
&\multispan2			&&\omit		&&	4.58012	&	1.97632	&	5.13971	&&	17.2723	&\cr \tablevert

&	-42652	&	61213	&&	15728825412	&&	2.7499	&	4.01931	&	1.60339	&&	7.63107	&\cr \drop
&\multispan2				&&\omit		&&	4.87457	&	2.03556	&	5.23786	&&	18.9611	&\cr \tablevert

&	-56788	&	5227	&&	15728825412	&&	1.78458	&	3.71019	&	4.6623	&&	12.2181	&\cr \drop
&\multispan2				&&\omit		&&	5.00997	&	4.30654	&	1.9168	&&	15.6404	&\cr \tablevert

&	-44489	&	60597	&&	17116856028	&&	2.94401	&	4.15224	&	1.63302	&&	8.44413	&\cr \drop
&\multispan2				&&\omit		&&	4.78675	&	2.01875	&	5.21508	&&	18.4752	&\cr \tablevert

&	-38564	&	62177	&&	17905442820	&&	2.32915	&	3.73889	&	1.56682	&&	6.10593	&\cr \drop
&\multispan2				&&\omit		&&	5.03011	&	2.06201	&	5.25529	&&	19.7369	&\cr \tablevert

&	-62628	&	27617	&&	18277035876	&&	1.59599	&	1.99402	&	3.54518	&&	5.18889	&\cr \drop
&\multispan2				&&\omit		&&	5.23878	&	5.12748	&	2.07445	&&	20.1064	&\cr \tablevert

&	-37761	&	62308	&&	18424313244	&&	2.24968	&	3.68906	&	1.56687	&&	5.85991	&\cr \drop
&\multispan2				&&\omit		&&	5.05495	&	2.06562	&	5.25394	&&	19.8438	&\cr \tablevert

&	-60689	&	16452	&&	19983541908	&&	1.62862	&	2.91726	&	4.13391	&&	8.32831	&\cr \drop
&\multispan2				&&\omit		&&	5.21882	&	4.79954	&	2.02126	&&	18.5473	&\cr \tablevert

&	-53411	&	55268	&&	20194538580	&&	3.91694	&	4.79004	&	1.83008	&&	13.368	&\cr \drop
&\multispan2				&&\omit		&&	4.12873	&	1.87726	&	4.91362	&&	14.5898	&\cr \tablevert

&	-62772	&	31033	&&	22250745252	&&	1.7132	&	1.75215	&	3.4649	&&	4.83898	&\cr \drop
&\multispan2				&&\omit		&&	5.20135	&	5.19072	&	2.07842	&&	20.2252	&\cr \tablevert

&	-60636	&	16253	&&	22886448612	&&	1.63115	&	2.93276	&	4.14453	&&	8.39525	&\cr \drop
&\multispan2				&&\omit		&&	5.21668	&	4.79216	&	2.01981	&&	18.5057	&\cr \tablevert

&	-52607	&	55964	&&	24231179244	&&	4.20956	&	1.89527	&	4.95837	&&	15.0652	&\cr \drop
&\multispan2				&&\omit		&&	3.82664	&	4.73501	&	1.8101	&&	12.8601	&\cr \tablevert

&	-60183	&	14636	&&	26167379412	&&	1.65248	&	3.05674	&	4.22936	&&	8.94485	&\cr \drop
&\multispan2				&&\omit		&&	5.19712	&	4.73034	&	2.0075	&&	18.153	&\cr \tablevert

&	-59739	&	13172	&&	27462020340	&&	1.67259	&	3.166	&	4.30375	&&	9.44933	&\cr \drop
&\multispan2				&&\omit		&&	5.17614	&	4.67153	&	1.99548	&&	17.8112	&\cr \tablevert

&	-37068	&	62407	&&	29569264236	&&	2.18249	&	3.64838	&	1.56959	&&	5.66433	&\cr \drop
&\multispan2				&&\omit		&&	5.07502	&	2.06835	&	5.25162	&&	19.9248	&\cr 
height2pt&\omit&\omit&&\omit&&\omit&\omit&\omit&&\omit&\cr
\tablerule
	}}

\vfill\eject
\bigskip
\hskip .15truein
\vbox{\topskip=0pt \offinterlineskip
\def\tablerule{\noalign{\hrule}}
\def\mystrut{\vbox to 9.5pt{\vss\vtop to 4.5pt{\vss\hbox to 0pt{}}}}
\def\tablevert{height2pt&\omit&\omit&&\omit&&\omit&\omit&\omit&&\omit&\cr 
	\tablerule
	height3pt&\omit&\omit&&\omit&&\omit&\omit&\omit&&\omit&\cr}
	
	\def\drop{height3pt&\omit&\omit&&\omit&&\omit&\omit&\omit&&\omit&\cr 
	}
	
\halign{ \vrule#\topskip=1em plus 2em\tabskip=10pt&
	\mystrut\hfil$#$\hfil& \hfil$#$\hfil&\vrule#&\hfil$#$\hfil&&\vrule#&\hfil$#$\hfil& \mystrut\hfil$#$\hfil& 
	\hfil$#$\hfil& \vrule#&\hfil$#$\hfil&\vrule#\tabskip=0pt\cr\tablerule
	height3pt&\omit&\omit&\omit&\omit&&\omit&\omit&\omit&\omit&\omit&\cr
	&\multispan4\hidewidth \mystrut Wallach Space\hidewidth&&\multispan5\hidewidth \mystrut Metric Parameters\hidewidth&\cr 
	height3pt&\omit&\omit&\omit&\omit&&\omit&\omit&\omit&\omit&\omit&\cr\tablerule
	height3pt&\omit&\omit&&\omit&&\omit&\omit&\omit&&\omit&\cr
	&k&l&&\lambda_h\mod 2^3\cdot 3\cdot \Gamma&&a&b&c&& \omit\hfil $\beta$ \hfil &\cr \tablevert

&	-35748	&	62561	&&	30559561404	&&	2.05927	&	3.5786	&	1.583	&&	5.33967	&\cr \drop
&\multispan2				&&\omit		&&	5.10987	&	2.0726	&	5.24438	&&	20.0512	&\cr \tablevert

&	-62739	&	29572	&&	30895359468	&&	1.64926	&	1.84639	&	3.48449	&&	4.9228	&\cr \drop
&\multispan2				&&\omit		&&	5.22013	&	5.16602	&	2.07751	&&	20.1979	&\cr \tablevert

&	-55339	&	2007	&&	34682602716	&&	1.8281	&	3.90803	&	4.78467	&&	13.3176	&\cr \drop
&\multispan2				&&\omit		&&	4.91823	&	4.13694	&	1.87909	&&	14.6378	&\cr
height2pt&\omit&\omit&&\omit&&\omit&\omit&\omit&&\omit&\cr
\tablerule
	}}

\refstyle{A}

\vfill\eject
\head  Figures\endhead

\input epsf

\nopagenumbers
\midinsert
\vskip 1.0in
\hskip 0in
\epsfbox{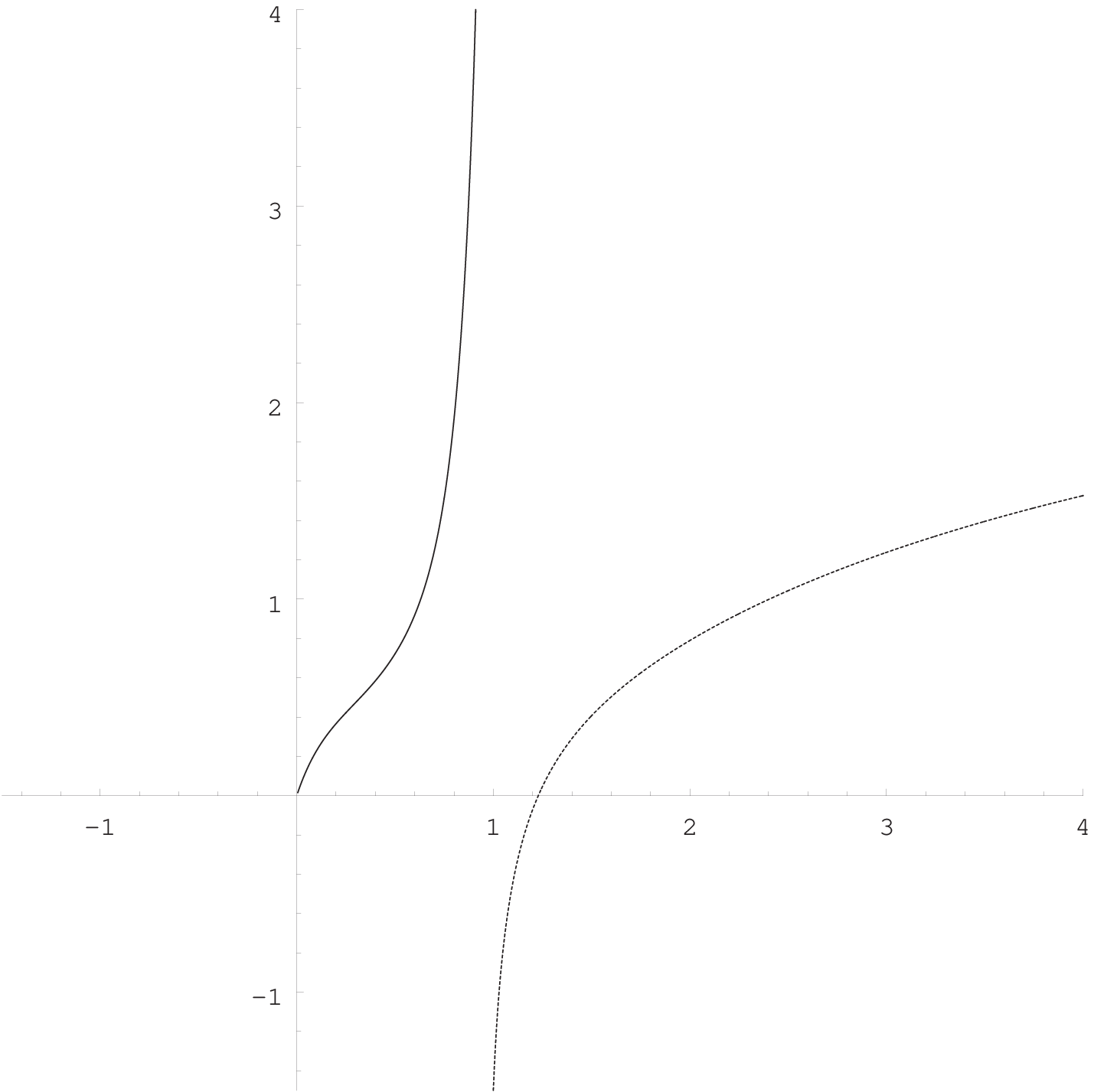}

\vskip 1.0in
\botcaption{Figure 1}
Curves of zeros for the Wallach space $M_{-42652,61213}$.\endcaption 
\endinsert

\vfill

\eject

\topinsert
\vskip 1.0in
\hskip .2in

\epsfbox{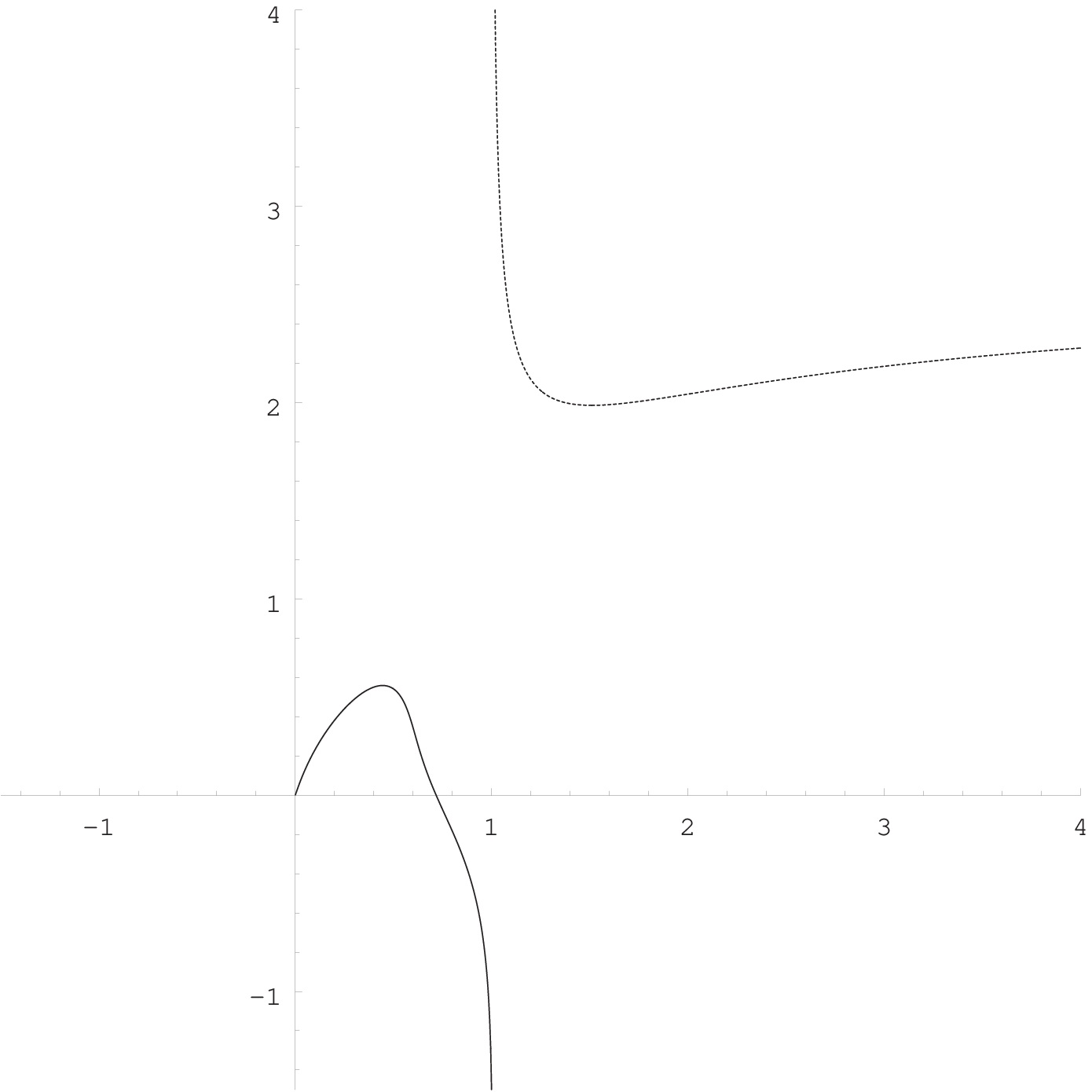}

\vskip 1.0in
\botcaption{Figure 2}
Curves of zeros for the Wallach space $M_{-42652,61213}$
under interchange  of $l$ and $(k+l)$.\endcaption

\endinsert

\vfill

\eject 

\topinsert
\vskip 1.0in
\hskip .2in

\epsfxsize=5in
\epsfysize=5in
\epsfbox{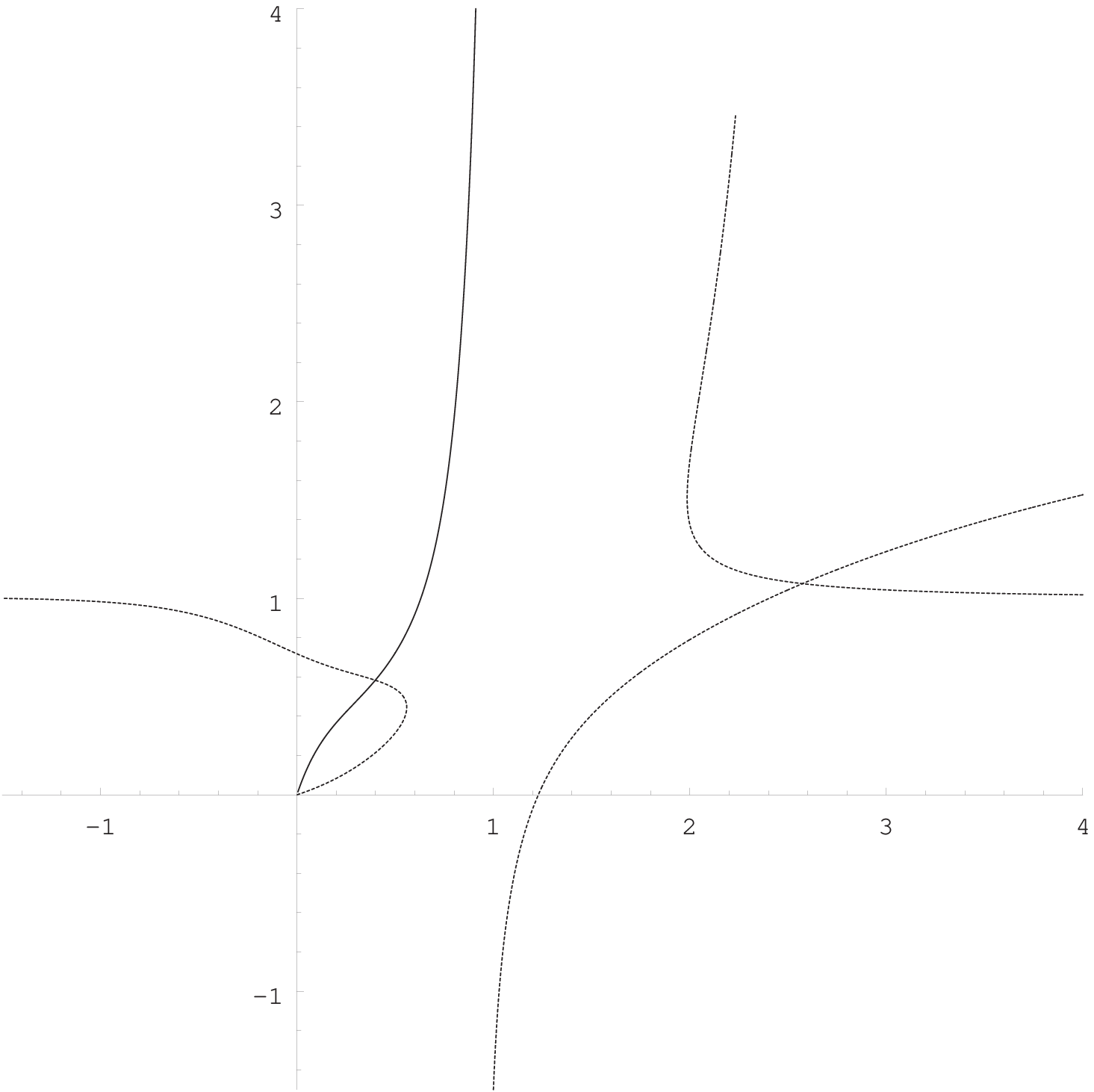}

\vskip 1.0in
\botcaption{Figure 3} The intersection of both curves of zeros for 
the Wallach space \hfil\break
$M_{-42652,61213}$. Points of intersection correspond to
Einstein metrics of nonegative curvature.\endcaption 
\endinsert

\vfill

\eject 
\topinsert
\vskip 1.0in
\hskip .2in

\epsfbox{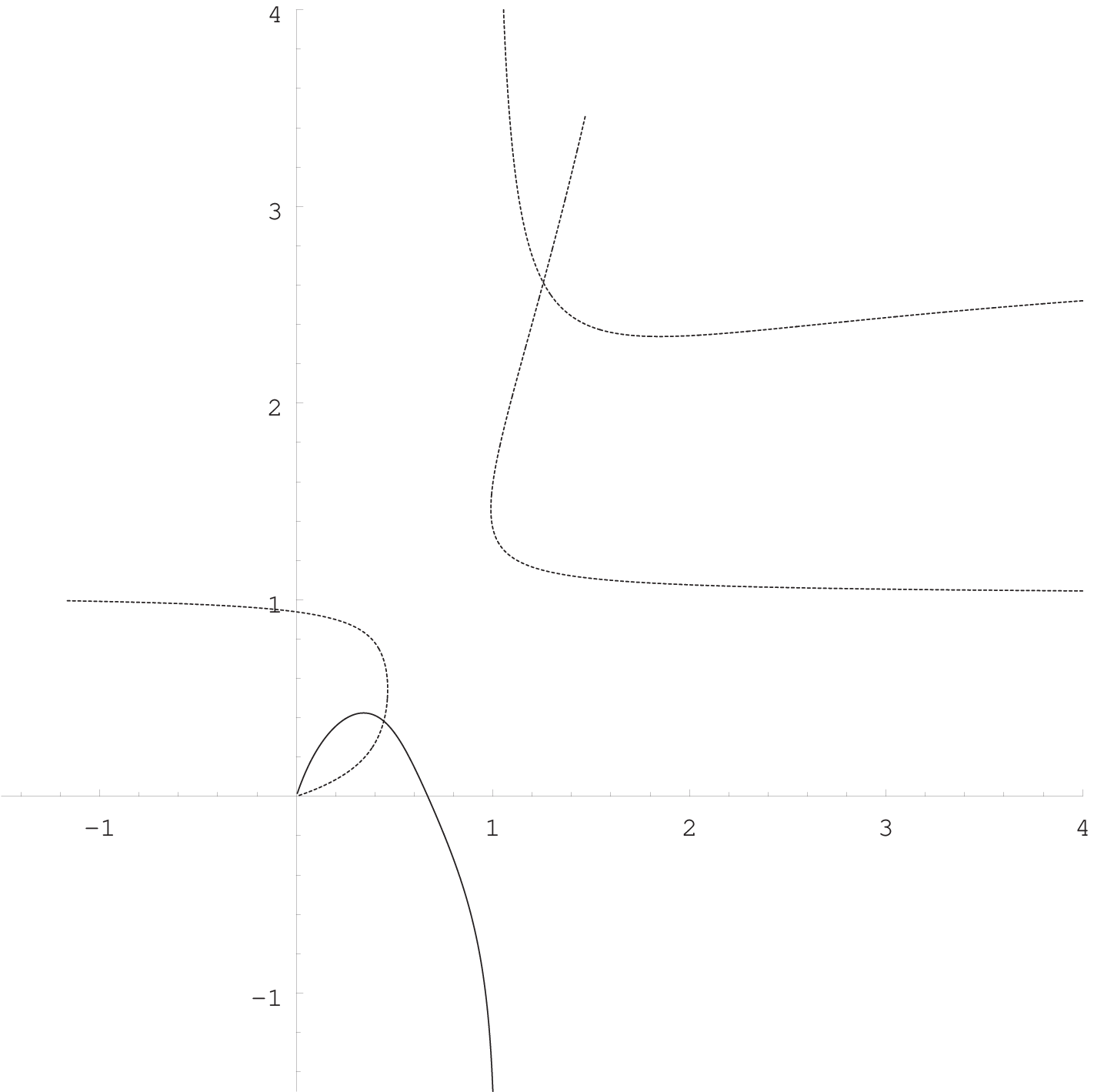}
\vskip 1.0in
\botcaption{Figure 4}
  The intersection of both curves of zeros for 
the Wallach space \hfil\break
$M_{-56788,5227}$. Note that again there are exactly two
intersections.\endcaption 
\endinsert

\vfill

\eject 

\topinsert
\vskip 1.0in
\hskip .2in

\epsfbox{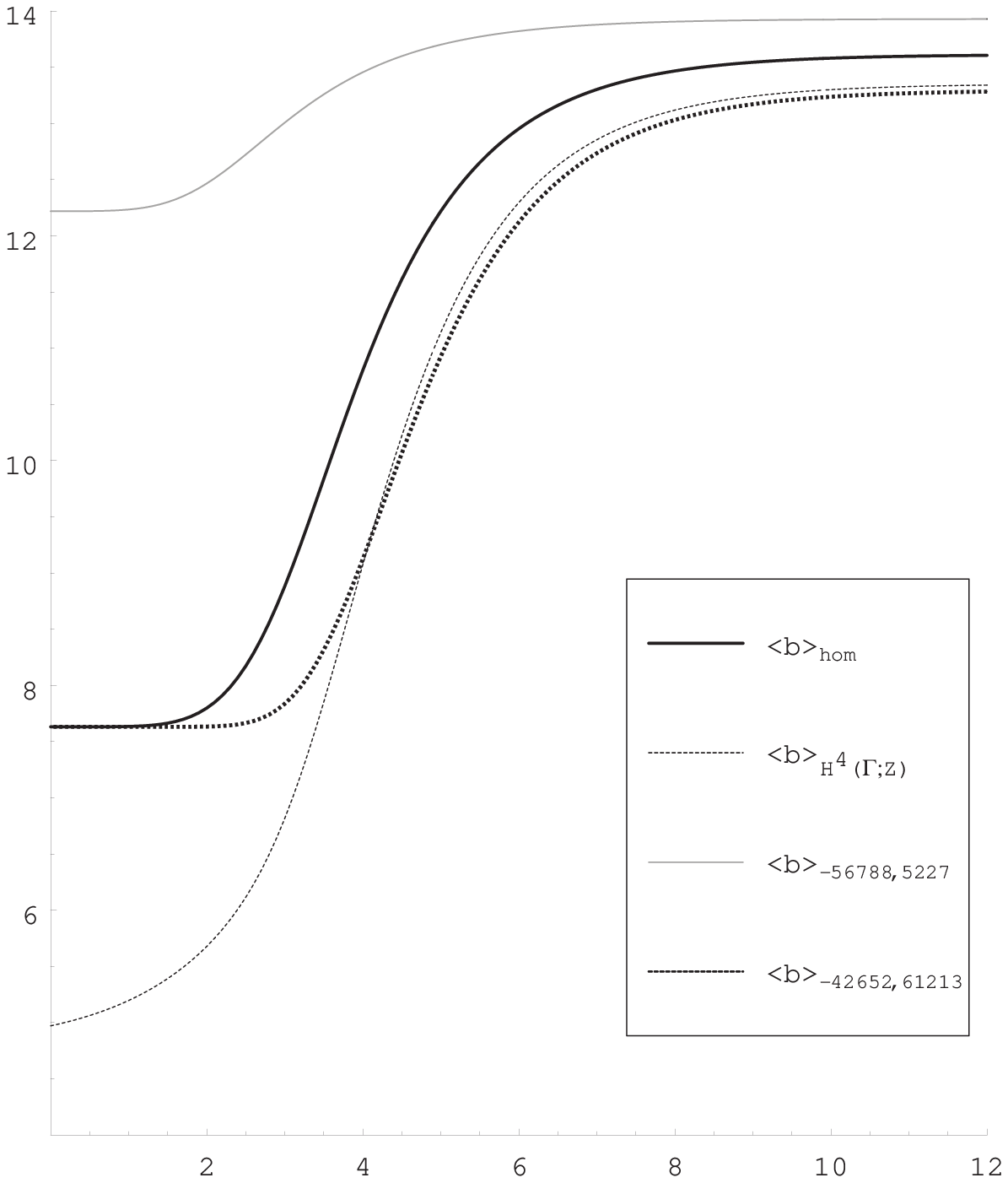}

\vskip 1.0in
\botcaption{Figure 5} The expectation value $<\beta>$ as a function of
cosmological constant for different sets of Wallach spaces.\endcaption 
\endinsert

\enddocument